\documentclass[a4paper,11pt]{article}
\pdfoutput=1
\usepackage{jinstpub}
\usepackage{graphicx}
\usepackage{booktabs}
\usepackage{siunitx}
\usepackage{xspace}
\newcommand{\Cherenkov}{\v{C}erenkov\xspace}

\title{\boldmath Time-over-threshold versus pulse-height energy reconstruction in a PbWO$_4$ calorimeter read out by SiPMs}

\author{S.~Carsi}
\author[a]{and E.~Vallazza}
\affiliation[a]{INFN Sezione di Milano Bicocca,\\Piazza della Scienza 3, 20126 Milano, Italy}
\emailAdd{stefano.carsi.uni@gmail.com}

\abstract{
The time over threshold (ToT) of a signal can be measured with a comparator and a time-to-digital converter, avoiding
the waveform digitizers normally used to read out calorimeter prototypes in test beams.
We assess how much the energy resolution degrades when the pulse height (PH), computed by the fast analysis
of the digitizers, is replaced by the ToT. The study uses the data collected in June 2024 at the CERN PS T9 beam line
with 2--6\,GeV electrons impinging on OREO, an electromagnetic calorimeter made of two $3\times3$ matrices of
PbWO$_4$ crystals read out by silicon photomultipliers (SiPMs) and sampled at 500\,MS/s, and 1--6\,GeV electrons on a lead-glass calorimeter
sampled at 100\,MS/s.
The ToT is emulated offline on the recorded waveforms, with and without interpolation of the threshold crossings, and it
is converted into an amplitude through a per-channel calibration curve, so that the same reconstruction chain can be
applied to both estimators. For OREO in random orientation the PH analysis gives
$\sigma/\mu = 3.6\%$ at 2\,GeV and $2.4\%$ at 6\,GeV. A single-threshold ToT worsens the resolution by 50--80\%,
because one threshold cannot cover the dynamic range of the calorimeter, while a dual-threshold ToT reaches
$\sigma/\mu = 3.9\%$ and $2.6\%$, i.e.\ 6--10\% worse than the PH. Extending the waveforms beyond the
\SI{1}{\micro\second} acquisition window with the measured exponential tail and noise shows that the window is not the
limiting factor: without it, a single threshold still worsens the resolution by about 60\%, while two thresholds give
the same result as before. An emulated shaper that reduces the decay time of the pulses to 50--100\,ns preserves this result with a ToT
of 0.3--0.5\,\si{\micro\second} instead of \SI{1.2}{\micro\second}. The conversion from ToT to amplitude can
also be calibrated with a scan of the two thresholds, without any digitizer, with a resolution comparable to the one
obtained with curves derived from the PH.
Interpolating the crossings brings no appreciable gain at 2\,ns sampling. For the lead glass, whose pulses are less than two samples wide at half maximum, the ToT cannot
be emulated reliably from the recorded waveforms. While re-deriving the PH reference we also found that the previously reported analysis of the
same data is consistent with a drift normalization that does not subtract the pedestals; subtracting them improves the
resolution by up to 1.6\,percentage points.}

\keywords{Calorimeters; Time over threshold; Front-end electronics for detector readout; Photon detectors for UV,
visible and IR photons (solid-state) (SiPMs)}

\begin{document}
\maketitle
\flushbottom

% =============================================================================================
\section{Introduction}
\label{sec:intro}

Calorimeter prototypes are usually read out in test beams with waveform digitizers: the amplitude of every channel,
or pulse height (PH), is obtained from the sampled waveform, either online or offline. For large-scale or
space-constrained applications a cheaper readout is attractive. A common choice is the time over threshold (ToT):
the signal is compared with a fixed threshold and the duration of the logic pulse is measured with a time-to-digital converter (TDC), which can be
integrated in an ASIC or implemented in an FPGA. The drawback is a strongly nonlinear relation between ToT and
amplitude: for a pulse with an exponential tail of time constant $\tau$ the ToT grows as $\tau\ln(A/\theta)$, where
$A$ is the amplitude and $\theta$ the threshold~\cite{Karpushkin2024}, so that large signals are compressed.
Several remedies have been proposed, mostly for positron emission tomography (PET), from dedicated shaping~\cite{PenaRodriguez2025} and dynamic
thresholds~\cite{Shimazoe:2012dtot} to the use of more than one threshold~\cite{Fujiwara:2010mltot,Gaudin:2020dualToT};
with two thresholds and a nonlinearity correction, the energy resolution of a PET detector was found to be at most
8--12\% worse than with an ADC~\cite{Gaudin:2020dualToT}. In calorimetry, the ToT is used to extend the dynamic range
of an ADC once the preamplifier saturates~\cite{Thienpont:2020wau}.

In this paper we use test-beam data to quantify how much the energy resolution of a SiPM-based homogeneous
calorimeter degrades when the PH is replaced by the ToT. We use OREO, a PbWO$_4$ calorimeter developed to exploit the
shower enhancement in oriented crystals~\cite{Bandiera2018,Soldani2025,Bandiera2025,Malagutti2024}. Since the
waveforms were recorded, the ToT can be emulated offline and compared event by event with the PH, keeping
everything else (event selection, equalization, calibration) identical. The study is restricted to the random
(non-channeling) orientation of the crystals. As a first step we re-derive the PH reference and compare it with the
analysis of the same data reported in ref.~\cite{Selmi2026}. We also examine the ToT of a lead-glass calorimeter
sampled at 100\,MS/s.

% =============================================================================================
\section{Experimental setup and data}
\label{sec:setup}

The data were collected in June 2024 at the CERN PS T9 beam line with electron beams of 1--6\,GeV and focused optics.
The setup consisted of a plastic scintillator used for the trigger, two silicon microstrip beam chambers~\cite{Prest2003}
(384 strips with \SI{242}{\micro\meter} readout pitch, $9.3\times9.3$\,cm$^2$ active area) placed 3.37\,m apart,
the OREO calorimeter, with its front face 25.75\,cm downstream of the second chamber, and a calorimeter of seven
lead-glass (LG) blocks ($10\times10$\,cm$^2$, 37\,cm long), 53\,cm downstream of the second chamber.
Two threshold \Cherenkov detectors upstream of the setup were used to tag electrons.

OREO consists of two $3\times3$ matrices of PbWO$_4$ crystals with a $2.5\times2.5$\,cm$^2$ cross section: the upstream
(UP) layer, 4.5\,cm long, can be oriented with respect to the beam, while the downstream (DOWN) layer, 10\,cm long, is
not oriented. Each crystal is read out by four Hamamatsu S13360-6050 SiPMs, biased at 53\,V, close to the breakdown
voltage, to avoid saturation, and AC-coupled to the digitizer. The SiPM signals were sampled by CAEN x730 digitizers
(14\,bit, 2\,V$_{\rm pp}$, 500\,MS/s or 2\,ns per sample, \SI{0.12}{\milli\volt} per ADC count), the LG photomultiplier signals and
the trigger scintillator by a CAEN V1724 (14\,bit, 100\,MS/s or 10\,ns per sample). The acquisition window contains
512 samples (\SI{1.02}{\micro\second} for OREO). The fast analysis of the digitizers provides, for every channel,
the baseline, the PH (difference between the baseline and the minimum of the negative signal) and the time of the
maximum.

\begin{figure}[htbp]
\centering
\includegraphics[width=\textwidth]{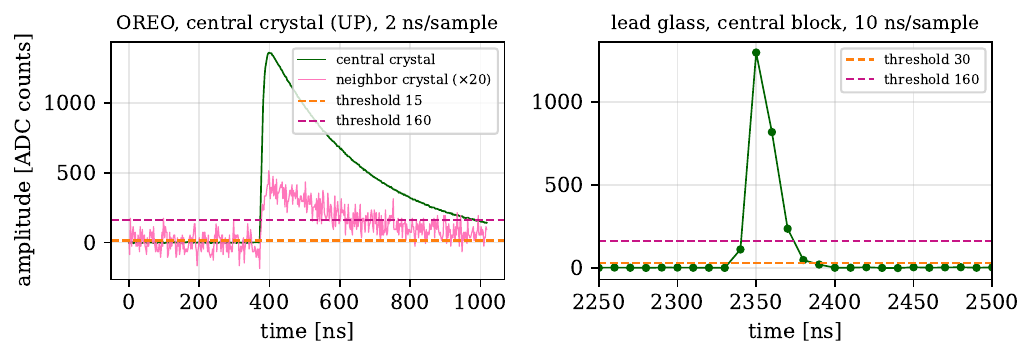}
\caption{Left: waveform of the central UP crystal of OREO for a 6\,GeV electron (green) and of a neighboring crystal
(pink, multiplied by 20). Right: waveform of the central LG block for a 6\,GeV electron (OREO out of the beam).
Dashed lines: two of the thresholds (ADC counts) used for the ToT emulation (they refer to the unscaled amplitude).}
\label{fig:wf}
\end{figure}

Figure~\ref{fig:wf} shows typical waveforms. The OREO pulse rises in about 20\,ns and decays slowly, with a tail that,
for the crystal hit by a 6\,GeV electron, still exceeds 100 ADC counts at the end of the window. The LG pulse has a
full width at half maximum (FWHM) of about 17\,ns, or 1.7 samples.

The runs used for the main results are listed in table~\ref{tab:runs}. The LG was characterized in dedicated runs with OREO moved out of
the beam line. All OREO runs were taken with the matrices in random orientation and the beam centered on the
central crystal; a 12\,GeV negative hadron beam, mostly made of pions, was used for the equalization with minimum ionizing
particles (MIPs).

\begin{table}[htbp]
\centering
\caption{Runs used in this analysis.}
\label{tab:runs}
\small
\begin{tabular}{llp{5.9cm}}
\toprule
purpose & beam & runs \\
\midrule
LG equalization & 1\,GeV e$^-$ on each block & 720381, 720384--720389 \\
LG calibration & 1--6\,GeV e$^-$ on central block & 720381, 720390--720394 \\
OREO MIP equalization & 12\,GeV negative hadrons ($\pi^-$) & 720395 \\
OREO energy scan & 2--6\,GeV e$^-$ & 2\,GeV: 720428, 720443; 3\,GeV: 720422, 720424, 720426;
4\,GeV: 720418, 720420; 5\,GeV: 720410, 720414, 720416; 6\,GeV: 720397, 720405 \\
ToT calibration curves & 12\,GeV $\pi^-$, 6\,GeV e$^-$ & 720395, 720465 \\
\bottomrule
\end{tabular}
\end{table}

% =============================================================================================
\section{Reference analysis with the pulse height}
\label{sec:ph}

\paragraph{Tracking and selection.}
The two chambers are aligned by fitting the distribution of the track angle with a Gaussian and shifting the second
chamber so that its mean is zero; the tracks are then extrapolated to the front faces of OREO and of the LG.
Electrons are tagged by requiring a signal in both \Cherenkov detectors. Events are further required to have the
impact point within a $1.5\times1.5$\,cm$^2$ box centered on the central crystal of OREO (a $3\times3$\,cm$^2$ box centered on
the central LG block for the LG runs), and the time of the maximum of the central channel(s) within a window around
the expected value. The crystal positions were determined from the occupancy maps measured with the pion beam.

\paragraph{Lead glass.}
The seven LG blocks are equalized with the 1\,GeV runs, each centered on one block, by fitting the PH peak with a
Gaussian. The equalization factors agree within 0.5--2\% with those of ref.~\cite{Selmi2026}. The equalized sum
of the seven blocks is linear with the beam energy, with residuals below 0.5\%, and the calibration line
$\mathrm{PH} = (193.1\pm0.4)\,\mathrm{ADC/GeV}\cdot E + (52.7\pm1.3)\,\mathrm{ADC}$ agrees within 1\% in slope and 9\% in intercept with those reported in
the logbook of the beam test ($191.6\pm0.2$, $53.3\pm0.4$) and in ref.~\cite{Selmi2026} ($195.1\pm0.5$, $48.5\pm2.7$).
The resolution, $\sigma/(\mu-q)$ with $q$ the intercept of the calibration line, decreases from 13.4\% at 1\,GeV to 8.3\% at
6\,GeV; a fit with $a/\sqrt{E}\oplus c$ gives $a=(11.6\pm1.0)\%\sqrt{\rm GeV}$ and $c=(6.4\pm0.5)\%$, with uncertainties
scaled by $\sqrt{\chi^2/\mathrm{ndf}}$ ($\chi^2/\mathrm{ndf}=9.3/4$). This constant term is dominated by the sampling: with 1.7 samples per
FWHM, the maximum sample depends on the phase between the pulse and the sampling clock. With the same events, the
integral of the waveform gives 4.5\% at 6\,GeV.

\begin{figure}[htbp]
\centering
\includegraphics[width=0.5\textwidth]{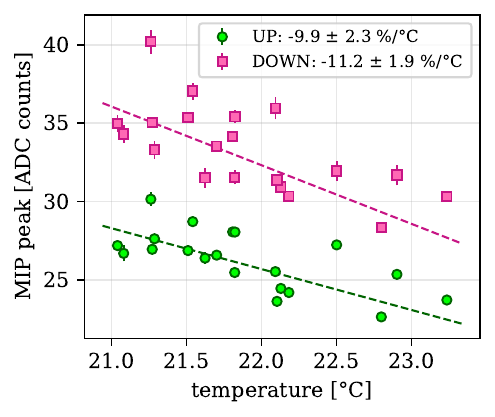}
\caption{MIP peak of the equalized layer sums versus the temperature recorded during each subrun (subruns with at
least 300 MIPs; the temperature was recorded from June 10 to June 17). Dashed: linear fits; the slopes, relative to the mean MIP peak, have uncertainties scaled
by $\sqrt{\chi^2/\mathrm{ndf}}$ to account for the scatter, possibly due to the thermal lag between the holder and
the SiPMs.}
\label{fig:temp}
\end{figure}

\begin{figure}[htbp]
\centering
\includegraphics[width=0.55\textwidth]{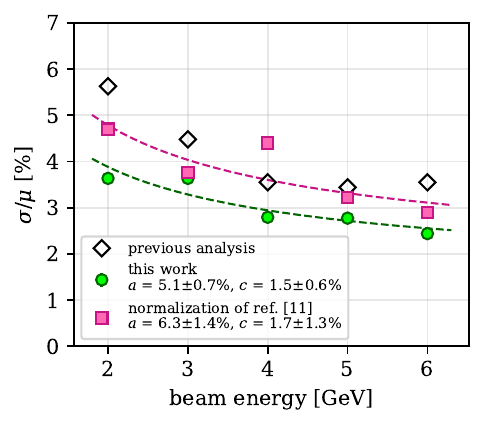}
\caption{Energy resolution of OREO (random orientation) with the PH: this work (green), the same analysis with the layer
sums normalized without pedestal subtraction (pink) and the values of ref.~\cite{Selmi2026} (open diamonds). Dashed lines:
fits with $a/\sqrt{E}\oplus c$, whose parameters are given in the legend with uncertainties scaled as in
table~\ref{tab:res}.}
\label{fig:phref}
\end{figure}

\paragraph{OREO.}
The reconstruction follows ref.~\cite{Selmi2026}:
\begin{enumerate}
\item the pedestal of every channel is the position of the noise peak of the PH distribution (about 8 ADC counts, since
the PH is the maximum of the noise when no signal is present);
\item the channels are equalized with MIPs: for every crystal the PH of the pions crossing its center without interacting is fitted with a
Landau function and the equalization factor is $Q_i=\mathrm{MPV}_{\rm CC}/\mathrm{MPV}_i$ within each layer, with MPV the most probable value
and CC the central crystal. The MPVs (17--38
ADC counts) agree within 1--5\% with the values reported in the logbook of the beam test;
\item the SiPM gain drifts with temperature during the data taking: the MIP peak of the layer sums increases by 14\% in
the UP and by 18\% in the DOWN layer between June 9 and June 15. The drift follows the temperature recorded by a
TMP37 sensor placed on the mechanical holder of OREO, a few cm from the crystals, and read out by an Arduino board
(figure~\ref{fig:temp}): the MIP peak decreases by $(9.9\pm2.3)\%$/$^\circ$C in the UP
and by $(11.2\pm1.9)\%$/$^\circ$C in the DOWN layer; given the temperature coefficient of 54\,mV/$^\circ$C of their operating
voltage~\cite{HamamatsuS13360}, this corresponds to SiPMs operated about 0.5\,V above breakdown. Each run is divided into
subruns of at most
50,000 events; in each subrun the equalized sum of each layer for MIP events (no \Cherenkov signal) is fitted with a Landau
function and the layer sum is divided by its MPV. At 2--4\,GeV the beam is almost purely made of electrons and some
subruns contain fewer than 300 MIPs; in this case the MPV is interpolated between the closest random-orientation runs at 6\,GeV,
which contain a large fraction of pions. Without the drift correction, the mean energy of run 720397 alone is 6\% lower and
its resolution 8\% worse;
\item each layer is calibrated by fitting the mean of its distribution versus the mean energy deposited in the layer
according to the Geant4 simulation of ref.~\cite{Selmi2026}, available at 2--6\,GeV, which limits the OREO
analysis to this range;
\item the energy is $E_{\rm OREO}=E_{\rm UP}+E_{\rm DOWN}$ (the LG is not included). Its distribution, which has a low-energy
tail due to leakage, is fitted with an ExpGaussExp function~\cite{Das2016} (referred to as Crystal Ball in
ref.~\cite{Selmi2026}), and the resolution is
$R=\sigma/\mu$, with $\sigma$ the width of the Gaussian core.
\end{enumerate}
The mean reconstructed energies, which correspond to the energy deposited in OREO according to the simulation, are
1.67, 2.56, 3.46, 4.25 and 5.06\,GeV at 2, 3, 4, 5 and 6\,GeV and agree within 2.5\% with
ref.~\cite{Selmi2026}. The resolution (green points in figure~\ref{fig:phref}) is
$R = 3.63, 3.64, 2.79, 2.77, 2.44\%$, with statistical uncertainties of 0.03--0.05\%.
A fit with $a/\sqrt{E}\oplus c$ gives $a=(5.1\pm0.7)\%\sqrt{\rm GeV}$ and $c=(1.5\pm0.6)\%$, with the uncertainties scaled by
$\sqrt{\chi^2/\mathrm{ndf}}$ ($\chi^2/\mathrm{ndf}=124/3$): the point-to-point fluctuations are dominated by run-dependent
systematic effects rather than by the statistics.

\paragraph{Comparison with the previous analysis.}
Our resolution is better than that of ref.~\cite{Selmi2026} by 0.7--2.0 percentage points, although the mean
energies agree within 2.5\%. In the layer sums of ref.~\cite{Selmi2026}, expressed in units of the MIP peak, the mean values are
about four times smaller than ours (e.g.\ 10.25 against 40.5 for the UP layer at 6\,GeV). The ratio is explained if the
nine pedestals ($\approx 9\times 8$ ADC counts) are not subtracted when the MIP peak of the layer sum is computed for the
drift normalization. When we repeat the analysis in this way we approximately reproduce the intermediate quantities of
ref.~\cite{Selmi2026}: the layer means in MIP units differ by 2--9\%, and the calibration constants (slope $m$, in MIP units per GeV, and
intercept $q$, in MIP units, of the layer calibration) are $m = 5.11$, $q = 2.21$ (UP) and $m = 3.34$, $q = 1.60$ (DOWN), against $5.21\pm0.22$, $1.88\pm0.31$ and $3.02\pm0.05$,
$1.64\pm0.16$. The resolution worsens to 4.70, 3.76, 4.40, 3.22 and 2.90\% (pink points in figure~\ref{fig:phref}),
because the pedestal part of the normalization does not drift and the correction is diluted. The residual differences
are compatible with the choice of the runs, which is not documented in ref.~\cite{Selmi2026}: with this normalization,
replacing the 6\,GeV runs with other random-orientation runs at the same energy changes the resolution at all energies, through
the layer calibration, by up to
1.6 percentage points, against at most 0.3 with our normalization at 2--5\,GeV. The runs at 6\,GeV taken after a
technical stop of the accelerator complex show a worse resolution (3.7\%) also with our normalization; we use the
runs recommended in the logbook (720397 and 720405) and quote the spread, 1.3 percentage points, as a systematic uncertainty on the 6\,GeV point; it is not included in
the fit with $a/\sqrt{E}\oplus c$.

% =============================================================================================
\section{Emulation of the time over threshold}
\label{sec:tot}

For every channel the baseline is computed on the first samples of the waveform and the contiguous region above
threshold that contains the pulse maximum is identified, for 16 thresholds between 3 and 640 ADC counts
(\SI{0.4}{}--\SI{78}{\milli\volt} for OREO; the MIP peak is at about 25 ADC counts, or \SI{3}{\milli\volt}). Two
flavors of ToT are computed:
\begin{itemize}
\item \emph{no interpolation}: number of samples above threshold times the sampling period, i.e.\ what a comparator
read out by a TDC running at the digitizer clock would measure. This flavor makes no assumption on the waveform between samples;
\item \emph{interpolation}: the threshold crossings are interpolated linearly between the two adjacent samples. This
approximates the ToT measured by a comparator and a fine TDC, as long as the pulse is well sampled
(section~\ref{sec:lg}).
\end{itemize}
If the signal is still above threshold at the end of the window, the ToT is truncated at the end of the window; a ToT
that ends less than 40\,ns before the end of the window is also treated as truncated.

\begin{figure}[htbp]
\centering
\includegraphics[width=\textwidth]{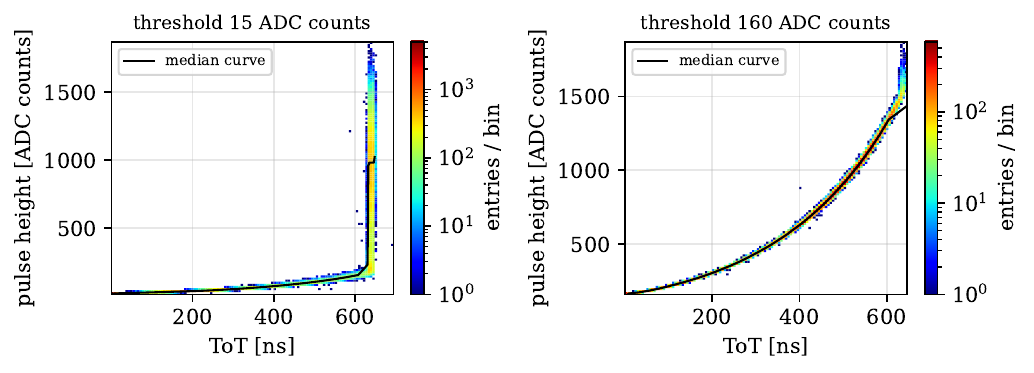}
\caption{PH versus interpolated ToT for the central UP crystal of OREO in run 720465 (6\,GeV e$^-$), for a single
threshold of 15 (left) and 160 (right) ADC counts, i.e.\ the two thresholds of the best dual-threshold configuration
(section~\ref{sec:results}); logarithmic color scale. Black: calibration curve (median PH in bins of ToT, runs 720395 and
720465). Left: the vertical band at 640\,ns is made of pulses truncated by the acquisition window. The pulses treated as
truncated (section~\ref{sec:tot}) form the last point of the curve, at the median of their PH; at 160 ADC counts this
flattens the curve above about 1400 ADC counts, for 5\% of the pulses of this run. If the last point includes only the
pulses actually truncated by the window, the energy resolution of OREO changes by less than 0.05 percentage points.}
\label{fig:totph}
\end{figure}

The ToT is converted into an amplitude by a per-channel calibration curve, the median PH as a function of the ToT,
built on runs not used for the resolution measurement (table~\ref{tab:runs}). Figure~\ref{fig:totph} shows the curves
for the central OREO crystal: at a threshold of 15 ADC counts (about 0.6 MIP) almost all electron signals are
truncated by the acquisition window and the curve saturates, whereas at 160 ADC counts the curve follows the data up to
about 1400 ADC counts. The fraction of pulses actually truncated by the window (not counting those that end in its last
40\,ns) in the central crystal in the 6\,GeV runs of the energy scan is
99.6\% at 15 ADC counts,
74\% at 80 ADC counts, 19\% at 120 ADC counts, 0.2\% at 160 ADC counts and zero above. This truncation is an artifact of the
\SI{1}{\micro\second} window and would not occur with a free-running comparator, which would however produce
ToTs of about \SI{1.2}{\micro\second} at a threshold of 10 ADC counts (section~\ref{sec:nowindow}).

The ToT-based amplitudes go through the same chain as the PH, with the same events, pedestals, equalization factors
and drift normalization; only the layer calibration and the resolution fit are redone. The comparison is thus
sensitive only to the amplitude estimator.

We also consider a dual-threshold scheme, which can be implemented with two comparators per channel connected to the
same signal, one with a low and one with a high threshold. Each comparator measures its own ToT, i.e.\ the time between
the crossing of its threshold on the leading edge and the crossing on the trailing edge, and each ToT is converted into
an amplitude with its own calibration curve. For every channel and event, the amplitude is taken from the
low-threshold ToT unless this is truncated by the acquisition window, in which case the high-threshold ToT is used.
The low threshold preserves the small signals of the outer crystals, the high one the large signals of the central
crystal. The scheme works like the dual-gain front ends used in calorimetry, where a high-gain channel measures the
small signals and a low-gain channel the large ones: here the two thresholds play the role of the two gains, and for
each channel and event only one of the two ToTs is used.

\begin{figure}[htbp]
\centering
\includegraphics[width=\textwidth]{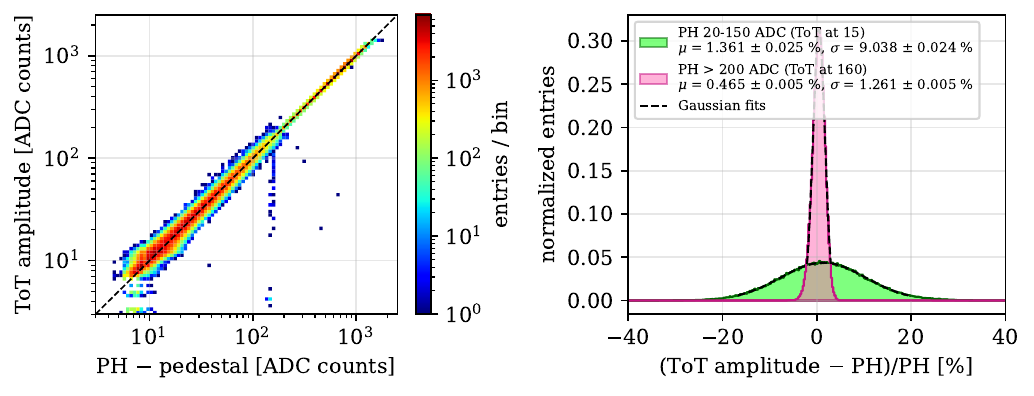}
\caption{Left: amplitude reconstructed from the dual-threshold ToT (15/160 ADC counts, interpolated) versus the PH,
for all OREO channels and the selected electrons at 2--6\,GeV; logarithmic color scale; dashed: diagonal. Right:
relative difference between the ToT amplitude and the PH for PH between 20 and 150 ADC counts (amplitude from the
low-threshold ToT) and above 200 ADC counts (high-threshold ToT), normalized to unit area; dashed: Gaussian fits with
uncertainties $\sqrt{N}$.}
\label{fig:dual}
\end{figure}

Figure~\ref{fig:dual} (left) compares the amplitude reconstructed in this way, for the configuration chosen in
section~\ref{sec:results} (15/160 ADC counts), with the PH, for all OREO channels and the selected electrons at
2--6\,GeV; figure~\ref{fig:dual} (right) shows the relative difference between the two in two ranges of PH. Between 20 and
150 ADC counts, where the low-threshold ToT is used for almost all signals, a Gaussian fit gives a mean of 1.4\% and a standard deviation of
9.0\%: the noise on the slowly decaying tail shifts the threshold crossing, and the PH itself fluctuates by about
3 ADC counts. Above 200 ADC counts, where the high-threshold ToT is used, the mean is 0.5\% and the standard deviation
1.3\%. Signals smaller than about 7 ADC counts above the pedestal do not cross the low threshold and are lost; with a
single threshold at 160 ADC counts, all signals below 160 ADC counts would be lost.

% =============================================================================================
\section{Results}
\label{sec:results}

\subsection{OREO}
\label{sec:oreo}

\paragraph{Figure of merit.}
For every ToT configuration we compute, at each energy, the ratio between the ToT and the PH resolution with two
estimators of the width: the core width $\sigma$ of the ExpGaussExp fit and $\sigma_{68}$, half of the central 68.3\%
interval. Each ratio is averaged over 2--6\,GeV and the larger of the two is taken as the degradation factor
\begin{equation}
\label{eq:F}
F = \max\left(\left\langle\frac{\sigma^{\rm ToT}}{\sigma^{\rm PH}}\right\rangle_{E},\;
\left\langle\frac{\sigma_{68}^{\rm ToT}}{\sigma_{68}^{\rm PH}}\right\rangle_{E}\right),
\end{equation}
where $\langle\cdot\rangle_E$ is the average over the beam energies of 2--6\,GeV; $F=1$ means no degradation.
The core width alone does not see tails, while $\sigma_{68}$ alone is less sensitive to a broadening of the core;
requiring both prevents one effect from hiding the other. When the core fit fails, $\sigma_{68}$ alone is used.
For the dual-threshold configurations, the five ratios that enter $F$ scatter by about 0.02 when the pedestals, equalization and drift correction are
derived from the PH, and by 0.1--0.15 when they are derived from the ToT; differences in $F$ below about 0.02 and 0.05,
respectively, are therefore not significant.

\begin{figure}[htbp]
\centering
\includegraphics[width=\textwidth]{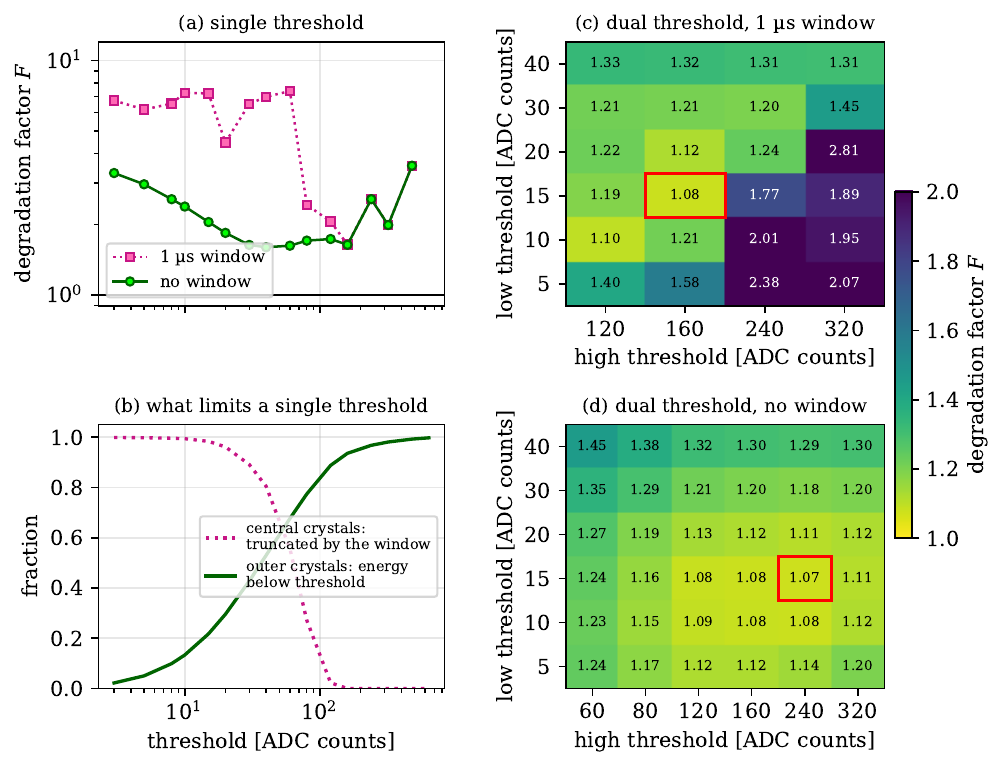}
\caption{Degradation factor $F$ (larger of the core-width and $\sigma_{68}$ ratios between ToT and PH, averaged over
2--6\,GeV) for the interpolated ToT. (a) Single threshold, with the \SI{1}{\micro\second} acquisition window and without
window (section~\ref{sec:nowindow}); missing points: spectra too distorted to be calibrated. (b) Fraction of the pulses
of the central crystals truncated by the window and fraction of the energy of the outer crystals carried by signals
below threshold, for electrons at 2--6\,GeV. (c), (d) Dual-threshold scheme with and without window; the red boxes mark
the chosen configurations.}
\label{fig:scan}
\end{figure}

\paragraph{Single threshold.}
Figure~\ref{fig:scan}a shows $F$ as a function of a single threshold, and figure~\ref{fig:scan}b two quantities that
explain its behavior: the fraction of pulses of the central crystals (UP and DOWN) that are still above threshold at the end
of the acquisition window, and the fraction of the energy of the outer crystals carried by signals below threshold.
Below about 60 ADC counts most pulses of the central crystals are truncated: their ToT saturates, it no longer depends
on the energy, and $F$ is 4.5--7.5. Above 120 ADC counts the central crystals are no longer truncated, but the signals of the
outer crystals fall below threshold. The fraction of the outer-crystal energy carried by signals below threshold grows
from 22\% at 15 ADC counts to 94\% at 160 ADC counts; since the outer crystals carry about 16\% of the energy measured
by OREO, this is 3.5\% and 15\% of the total, respectively. The best single threshold, 160 ADC counts, gives $F=1.6$. At higher thresholds the signals of the central
crystals at the lowest energies approach the threshold and $F$ increases again.

\paragraph{Dual threshold.}
The dual-threshold scheme removes this trade-off. Figure~\ref{fig:scan}c shows $F$ for low thresholds of 5--40
and high thresholds of 120--320 ADC counts. Two effects determine the optimum.
\begin{itemize}
\item The low threshold must be well above the noise. On the exponential tail of the pulse ($\tau\approx270$\,ns) a
noise fluctuation $\sigma_n$ shifts the threshold crossing by about $\sigma_n\tau/\theta$, so that the relative error on
the reconstructed amplitude scales as $\sigma_n/\theta$, with $\sigma_n\approx3$ ADC counts. On the other hand, a low
threshold above 20 ADC counts loses the smallest signals of the outer crystals.
\item The high threshold must not be too high. With a low threshold of 15 ADC counts the pulses larger than about
130 ADC counts are treated as truncated, since $15\,\exp(580\,\mathrm{ns}/270\,\mathrm{ns})\approx130$, where 580\,ns is the
time between the maximum and 40\,ns before the end of the window. Pulses with an amplitude between this value and the high
threshold have a truncated low-threshold ToT but do not cross the high threshold, so that their amplitude is lost: they
are assigned a value close to the pedestal of the channel, i.e.\ about zero after pedestal subtraction. This gap is small with a high threshold of
160 ADC counts and grows at 240 and 320 ADC counts.
\end{itemize}
The best configuration is 15/160 ADC counts, i.e.\ about 0.6 and 6 MIPs (1.8 and \SI{20}{\milli\volt}), with $F=1.08$; the combinations with a low threshold of 10--20 and a high
threshold of 120--160 ADC counts give $F=1.08$--1.22. The choice of the figure of merit matters: the core width alone
would select 15/240 ADC counts, whose spectrum at 2\,GeV has a tail that makes $\sigma_{68}$ four times larger than with
the PH, because at this energy the central crystal often falls in the gap between the two thresholds; $\sigma_{68}$ alone
would select 10/120 ADC counts, which broadens the core by 19\% at 4\,GeV.

\begin{figure}[htbp]
\centering
\includegraphics[width=\textwidth]{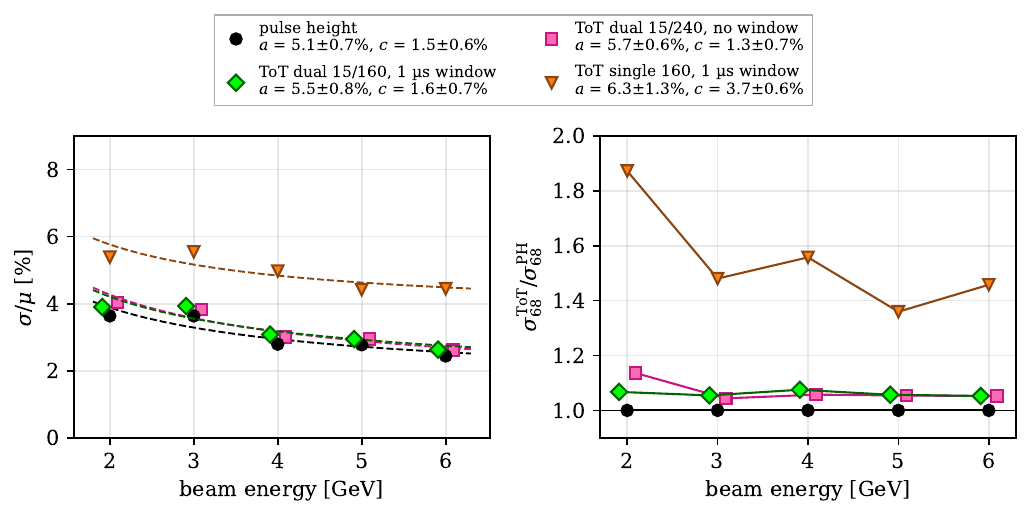}
\caption{Left: energy resolution of OREO (random orientation) with the PH, with the best single- and dual-threshold ToT
with the \SI{1}{\micro\second} window, and with the best dual-threshold ToT without window (section~\ref{sec:nowindow});
interpolated ToT; dashed lines: fits with $a/\sqrt{E}\oplus c$, whose parameters are given in the legend with
uncertainties scaled as in table~\ref{tab:res}. Right: ratio between $\sigma_{68}$ of the ToT and of the PH.
The ToT without interpolation gives the same resolution within 0.1 percentage points (table~\ref{tab:res}). The
dual-threshold points are shifted horizontally by $\pm0.09$\,GeV for readability.}
\label{fig:tot}
\end{figure}

\paragraph{Energy resolution.}
The resolution obtained with the best single and dual thresholds is compared with the PH in figure~\ref{fig:tot} and
in table~\ref{tab:res}. With a single threshold the core width is 50--80\% larger than with the PH and the constant
term grows from 1.5\% to 3.7\%. With two thresholds the core width is 6--10\% and $\sigma_{68}$ 5--8\% larger than with
the PH at all energies; the stochastic and constant terms (5.5--5.7\% and 1.5--1.6\%) are compatible with those of the PH
within the uncertainties. At 2\,ns sampling the interpolation of the crossings makes no appreciable difference, since the
relevant ToTs span tens to hundreds of samples.

\paragraph{Calibrations from the ToT.}
In the comparison above the pedestals, the equalization and the drift correction are taken from the PH chain. We
repeated the analysis deriving all of them, together with the time cuts, from the dual-threshold ToT alone
(table~\ref{tab:res}, $^\dagger$): the MIP peak is clearly visible with the ToT, the equalization factors agree with the
PH ones within 6\%, and the average degradation factor is almost unchanged ($F=1.09$), but the resolution fluctuates
more from one energy to the next, from 17\% better at 3\,GeV to 14--17\% worse at 5 and 6\,GeV than with the PH-based
calibration. The drift correction is the most likely cause: the MIP peak measured with the ToT follows the PH one, but
their ratio changes by $\pm5\%$ from one subrun to the next, so that runs taken at different temperatures no longer
overlap.

\begin{table}[htbp]
\centering
\caption{OREO energy resolution $\sigma/\mu$ (\%) in random orientation, with $\sigma$ the core width of the ExpGaussExp
fit. Statistical uncertainties are 0.03--0.06\%. The fit parameters of $a/\sqrt{E}\oplus c$ have uncertainties scaled by
$\sqrt{\chi^2/\mathrm{ndf}}$. $F$: degradation factor with respect to the PH (see text). Thresholds in ADC counts
(1 ADC count = \SI{0.12}{\milli\volt}). $^\dagger$Pedestals, equalization, drift correction and time cuts
derived from the ToT instead of the PH (section~\ref{sec:oreo}). $^\ddagger$ToT-to-amplitude conversion self-calibrated with a threshold scan and all calibrations derived from the ToT (section~\ref{sec:nodigitizer}). For ref.~\cite{Selmi2026}, the $a$ and $c$
columns give the fit that also includes data taken at the CERN SPS. $^*$Systematic uncertainty from the choice of
the 6\,GeV runs: $+1.3$ percentage points (section~\ref{sec:ph}).}
\label{tab:res}
\small
\setlength{\tabcolsep}{3.5pt}
\begin{tabular}{lcccccccc}
\toprule
estimator & 2\,GeV & 3\,GeV & 4\,GeV & 5\,GeV & 6\,GeV & $a$ [\%$\sqrt{\rm GeV}$] & $c$ [\%] & $F$ \\
\midrule
PH (this work)                      & 3.63 & 3.64 & 2.79 & 2.77 & 2.44$^*$ & $5.1\pm0.7$ & $1.5\pm0.6$ & -- \\
PH, normalization of ref.~\cite{Selmi2026} & 4.70 & 3.76 & 4.40 & 3.22 & 2.90 & $6.3\pm1.4$ & $1.7\pm1.3$ & -- \\
ref.~\cite{Selmi2026}               & 5.63 & 4.48 & 3.55 & 3.44 & 3.55 & \multicolumn{2}{c}{(T9+SPS fit: 5.95, 2.60)} & -- \\
\midrule
ToT single 160, interp.             & 5.39 & 5.54 & 4.97 & 4.42 & 4.45 & $6.3\pm1.3$ & $3.7\pm0.6$ & 1.64 \\
ToT single 160, no interp.          & 5.43 & 5.57 & 5.03 & 4.48 & 4.46 & $6.3\pm1.2$ & $3.7\pm0.6$ & 1.65 \\
ToT dual 15/160, interp.            & 3.90 & 3.93 & 3.08 & 2.94 & 2.63 & $5.5\pm0.8$ & $1.6\pm0.7$ & 1.08 \\
ToT dual 15/160, no interp.         & 3.94 & 4.02 & 3.07 & 3.00 & 2.64 & $5.7\pm0.9$ & $1.5\pm0.8$ & 1.09 \\
ToT dual 15/160, ToT-only calib.$^\dagger$ & 3.91 & 3.27 & 2.81 & 3.35 & 3.09 & $3.9\pm1.0$ & $2.5\pm0.5$ & 1.09 \\
\midrule
\multicolumn{9}{l}{without acquisition window (section~\ref{sec:nowindow})} \\
ToT single 40, interp.              & 5.78 & 5.56 & 4.68 & 4.31 & 4.00 & $7.8\pm0.8$ & $2.6\pm0.6$ & 1.60 \\
ToT dual 15/240, interp.            & 4.04 & 3.83 & 3.01 & 2.95 & 2.61 & $5.7\pm0.6$ & $1.3\pm0.7$ & 1.07 \\
ToT dual 15/240, no interp.         & 4.07 & 3.86 & 3.02 & 2.97 & 2.62 & $5.7\pm0.6$ & $1.4\pm0.7$ & 1.08 \\
ToT dual 15/240, no digitizer$^\ddagger$ & 4.15 & 3.28 & 2.76 & 2.94 & 2.63 & $5.2\pm0.6$ & $1.6\pm0.5$ & 1.03 \\
\bottomrule
\end{tabular}
\end{table}

\subsection{OREO without acquisition window}
\label{sec:nowindow}

\begin{figure}[htbp]
\centering
\includegraphics[width=\textwidth]{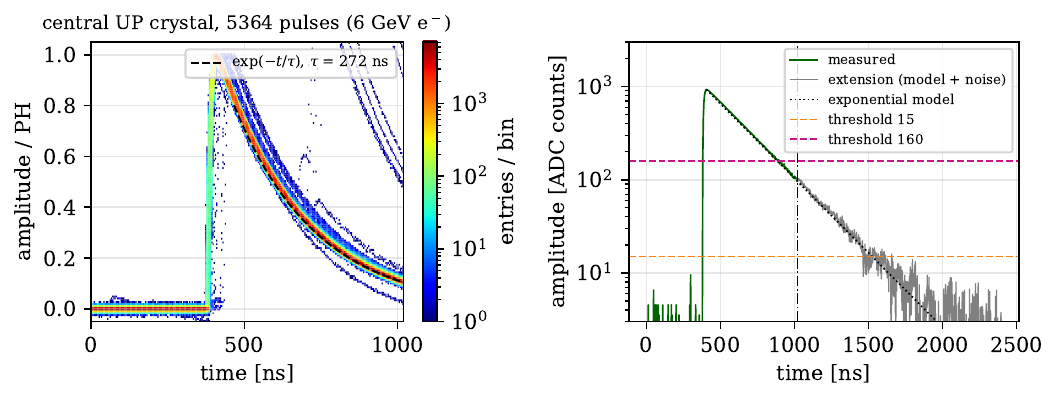}
\caption{Left: persistence plot of the waveforms of the central UP crystal for 6\,GeV electrons (run 720465), normalized
to their PH; logarithmic color scale; dashed: exponential with the fitted decay constant. Right: example of a waveform
of the same crystal extended beyond the end of the acquisition window (dash-dotted line) with the exponential model
(dotted) and synthesized noise; dashed: thresholds of 15 and 160 ADC counts.}
\label{fig:persist}
\end{figure}

A free-running comparator is not limited by an acquisition window. To emulate this case we extend each OREO waveform
beyond the window. The tail of the SiPM pulses is a single exponential (figure~\ref{fig:persist}, left): averaged over
pulses of 300--1500 ADC counts, the decay constant is 270--272\,ns for the channels read out by the first digitizer and
274--283\,ns for the others, independent of the amplitude, with residuals below 0.15\% of the pulse height. After the
last sample each waveform is continued up to \SI{4.1}{\micro\second} with $A\exp[-(t-t_{\rm max})/\tau]$, where $\tau$ is
fixed for each channel and $A$ is fitted on the measured tail, plus Gaussian noise generated with the power spectrum
measured on the pre-trigger samples of the same channel (rms of 2.5--3.2 ADC counts), as in figure~\ref{fig:persist}
(right). The noise must be included: on a slowly decaying tail it dominates the fluctuations of the trailing
threshold crossing. The ToT is then computed on the extended waveform with the same definitions as before.

The procedure was validated on pulses whose ToT ends inside the window. The waveform is cut 120\,ns before the measured
trailing crossing and extended as above, and the ToT distribution in bins of PH is compared with the measured one,
for thresholds of 15, 30 and 60 ADC counts. Away from the edges of the selection, the medians agree within 3\,ns
(below 1\% of the ToT) and the widths within 10\%, the emulated ones being slightly larger. The procedure does not
reproduce a weak slow noise component, about 5\% of the noise power, which stays correlated up to the longest delay we
measured, about 80\,ns, and
cannot be characterized on the short pre-trigger region. It also ignores the undershoot due to the AC coupling of the
SiPMs: a fit of the average tail of large pulses with an exponential and a constant limits the undershoot to 0.1\% of
the PH for the central UP crystal and 0.7\% for the central DOWN one, at most about 10 ADC counts for the largest
pulses. This could shorten the low-threshold ToT of the largest pulses: it does not affect the dual-threshold scheme,
which uses the high threshold for them, but it could worsen the single-threshold results at low thresholds.

Figure~\ref{fig:scan}a compares the degradation factor for a single threshold with and without window. Without
window the low thresholds no longer saturate and $F$ drops from 6.2--7.3 to 2--3.3 below 20 ADC counts, but it never
falls below 1.6: at low thresholds the noise on the tail degrades the amplitude of the central crystals, at high
thresholds the signals of the outer crystals are lost, as with the window. The best single threshold, 40 ADC counts,
gives $F=1.60$, close to the value obtained with the window (1.64 at 160 ADC counts). With two thresholds the
high-threshold ToT is used whenever the signal crosses the high threshold. The gap between the two thresholds
disappears, and $F$ is 1.07--1.09 for low thresholds of 10--15 and high thresholds of 120--240 ADC counts
(figure~\ref{fig:scan}d), with the best value, 1.07, at 15/240 ADC counts. The resolution
(figure~\ref{fig:tot} and table~\ref{tab:res}) differs from the one obtained with the window by at most
0.15 percentage points. The acquisition window is therefore not what limits the ToT: a single comparator is limited by
the dynamic range of the calorimeter, while two comparators reproduce the PH resolution on average within about 8\%, with or
without window. Without window the choice of the thresholds is much less critical.

Two variants were also tested without window. A single comparator with hysteresis, which switches on when the signal
rises above $\theta_{\rm high}$ and off when it falls below $\theta_{\rm low}\le\theta_{\rm high}$, gives its best result,
$F=1.62$, for $\theta_{\rm low}=\theta_{\rm high}=60$ ADC counts ($\theta_{\rm high}$ of 15--240 and $\theta_{\rm low}$ of 5--60 ADC
counts were tested), i.e.\ without hysteresis: lowering $\theta_{\rm low}$
moves the trailing crossing to where the noise is relatively larger, and $F$ grows up to 3.6. Using three or four
comparators per channel, with the amplitude taken from the highest threshold crossed, reduces $F$ only from 1.07 to
1.04--1.06 (e.g.\ 1.04 for 10/30/80/240 ADC counts).

\paragraph{Shorter pulses.}
Without window the ToT of the central crystal at a threshold of 10 ADC counts lasts about \SI{1.2}{\micro\second} at 6\,GeV, which
limits the rate per channel. The pulses can be shortened by an analog shaper before the comparators. We emulated it on
the recorded waveforms with a pole-zero cancellation, which turns the exponential tail with time constant $\tau_1$ (270--283\,ns) into
a shorter one with time constant $\tau_2$, followed by an RC low-pass filter with a time constant of 10\,ns, applied as a
digital filter to signal and noise. The gain is normalized so that the thresholds remain in PH-equivalent ADC counts.
The shaped pulses end inside the acquisition window, so that no extrapolation is needed, and the high-threshold ToT is
used whenever the signal crosses the high threshold. Table~\ref{tab:shaper} summarizes the results. With
$\tau_2=50$--100\,ns the dual-threshold ToT gives $F=1.05$--1.07, as without shaper, with a ToT 2.4--4.4 times shorter.
The low-pass filter is essential: without it the pole-zero cancellation amplifies the high-frequency noise and $F$
rises to 1.12. A single threshold improves to $F\approx1.43$, but it remains far from two thresholds. The emulation also
filters the noise of the digitizer, which an analog comparator would not see, so that these numbers are indicative.

\begin{table}[htbp]
\centering
\caption{Dual- and single-threshold ToT after an emulated shaper (pole-zero cancellation $\tau_1\to\tau_2$ and RC
low-pass filter), compared with the window-less ToT without shaper. $F$: degradation factor with respect to the PH,
for the best thresholds (ADC counts). Last column: median ToT of the central UP crystal at 6\,GeV at a threshold of
10 ADC counts.}
\label{tab:shaper}
\small
\setlength{\tabcolsep}{4pt}
\begin{tabular}{lccccc}
\toprule
shaper & low-pass & dual thresholds & $F$ dual & $F$ single (threshold) & ToT at 10 ADC counts \\
\midrule
none (no window)          & --     & 15/240 & 1.075 & 1.60 (40) & \SI{1.16}{\micro\second} \\
$\tau_2=25$\,ns           & 10\,ns & 10/160 & 1.097 & 1.49 (40) & 154\,ns \\
$\tau_2=50$\,ns           & 10\,ns & 10/120 & 1.065 & 1.44 (30) & 265\,ns \\
$\tau_2=100$\,ns          & 10\,ns & 10/160 & 1.054 & 1.43 (30) & 486\,ns \\
$\tau_2=150$\,ns$^*$      & 10\,ns & 10/120 & 1.047 & 1.44 (30) & 638\,ns \\
$\tau_2=50$\,ns           & --     & 15/240 & 1.122 & 1.69 (160) & 237\,ns \\
\bottomrule
\multicolumn{6}{l}{\footnotesize $^*$ the low-threshold ToT of the largest pulses reaches the end of the window.}
\end{tabular}
\end{table}

\subsection{Calibration without a digitizer}
\label{sec:nodigitizer}

So far the ToT has been converted into an amplitude with curves built from the PH, i.e.\ with a digitizer. A readout
without a digitizer must derive this conversion from the ToT itself. Inverting the logarithmic
relation of section~\ref{sec:intro}, $A=\theta\exp(\mathrm{ToT}/\tau)$, does not work: the ToT also includes the rise
of the pulse and the faster decay near the maximum, and the exponential amplifies these deviations, so that the
largest signals are overestimated by a factor of 2.5--10.

The pulse shape, on the other hand, depends only weakly on the amplitude. For the central UP crystal, the average pulse
normalized to its PH is 206--216\,ns wide at half maximum and 470--490\,ns wide at 20\% of the maximum for PH between
60 and 1600 ADC counts. The ToT of a pulse of amplitude $A$ at a threshold $\theta$ then depends only on $\theta/A$,
and $\ln(\theta/A)=g(\mathrm{ToT})$ with a single function $g$ per channel. Two thresholds crossed by the same pulse
give
\begin{equation}
g(\mathrm{ToT}_j)-g(\mathrm{ToT}_k)=\ln(\theta_j/\theta_k),
\label{eq:selfcal}
\end{equation}
which does not contain $A$: $g$ can be measured by scanning the settings of the two comparators, with no reference
amplitude. The condition $g(0)=0$ (the ToT vanishes when the threshold reaches the maximum) fixes the absolute scale,
which is in any case absorbed by the MIP equalization, and the amplitude is $A=\theta\exp[-g(\mathrm{ToT})]$.

\begin{figure}[htbp]
\centering
\includegraphics[width=\textwidth]{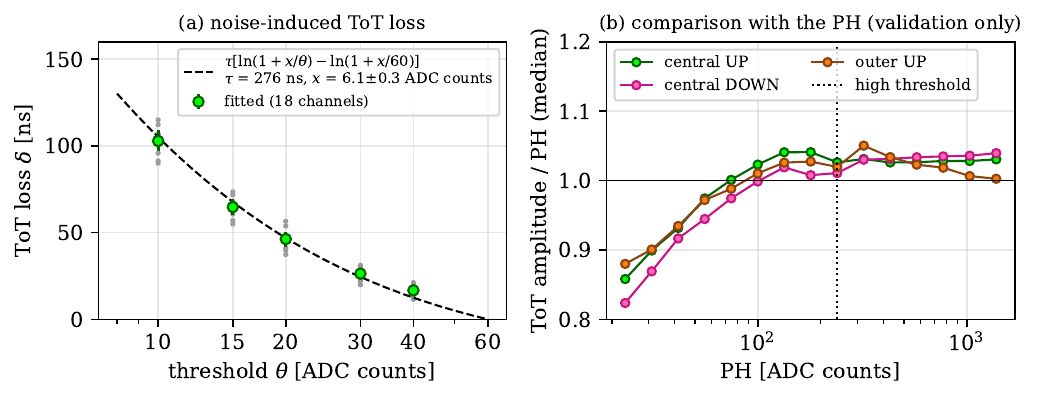}
\caption{Calibration without a digitizer. Left: ToT loss $\delta$ at the threshold crossing, fitted together with the
ToT-to-amplitude function; gray: single channels; markers: mean and standard deviation over the 18 channels; dashed:
$\tau[\ln(1+x/\theta)-\ln(1+x/60)]$ with $\tau$ the average decay constant of the tail; the fitted $\delta$ are relative
to the one at 60 ADC counts. Right: median ratio between the
self-calibrated amplitude (dual threshold 15/240 ADC counts) and the PH, runs 720395 and 720465, for the two central
crystals and one outer crystal; the PH is used only for this check.}
\label{fig:digifree}
\end{figure}

At low thresholds the noise breaks this relation. When the tail approaches the threshold, a negative fluctuation ends
the contiguous region above threshold early, and the ToT is shortened by $\delta(\theta)$. On the exponential tail the
slope at the crossing is $\theta/\tau$ for any amplitude, so $\delta$ depends on the threshold and not on $A$. We
therefore replace $g(\mathrm{ToT}_j)$ with $g(\mathrm{ToT}_j+\delta_j)$ in eq.~(\ref{eq:selfcal}) and fit one
$\delta_j$ per threshold together with $g$. On the exponential tail, shifts $\delta_j$ that vary linearly with
$\ln\theta_j$ are equivalent to a change of the slope of $g$, and only the deviations from this trend can be measured.
We set $\delta_j=0$ for $\theta\ge60$ ADC counts, so that the fitted $\delta_j$ are relative to the shift at
60 ADC counts, and neglect the variation of $\delta$ among these higher thresholds; the thresholds below 10 ADC counts,
which lie in the noise, are not used. The fitted $\delta_j$ are the same in all channels, with a standard deviation of
2--6\,ns, and follow $\delta(\theta)-\delta(60)$, with $\delta(\theta)=\tau\ln(1+x/\theta)$ the time the exponential
tail takes to decrease from $\theta+x$ to $\theta$ and $x=6.1\pm0.3$ ADC counts, about twice the noise rms
(figure~\ref{fig:digifree}a): the trailing crossing becomes unreliable once the signal is less than about two noise rms
above the threshold.

The function $g$ is described by its values at 110 points in ToT, spaced by 2\,ns up to 40\,ns and logarithmically up
to \SI{4.1}{\micro\second}, with linear interpolation in between and a small regularization term that keeps it smooth where the data are
sparse. It is determined for
each channel with a least-squares fit on runs 720395 and 720465, using all pairs of the 13 thresholds between 10 and 640 ADC
counts. To mimic a threshold scan, each pair of thresholds uses a separate subset of the events, as if each setting
were taken in a different run. Pairs of distant thresholds are needed: with adjacent thresholds only, $g$ is poorly
constrained at long ToT in the outer crystals, which see few large signals, and their low-threshold amplitude is
15--45\% too low near the switch to the high threshold. This can be checked without a digitizer on the signals that
cross both thresholds: with all pairs, the ratio of the low- and high-threshold amplitudes is 0.95--1.01 in all
channels. The PH, not used in the procedure, confirms the result (figure~\ref{fig:digifree}b): above 60 ADC counts the
ToT amplitude agrees with the PH within 5\%; below, the PH is itself biased upward by the noise, since it is the
maximum of the sampled waveform.

The rest of the chain uses no information from the digitizer either: the pedestal is zero, since a signal below the
low threshold gives no ToT, the equalization and the drift correction use the MIP peak of the ToT amplitudes, and the
time cuts use the leading edge at the low threshold. The \Cherenkov tag and the tracking are unchanged, since they
need only a discriminator and the silicon detectors. With the dual threshold at 15/240 ADC counts and no window, the
resolution is 4.15, 3.28, 2.76, 2.94 and 2.63\% from 2 to 6\,GeV (table~\ref{tab:res}), and $F=1.03\pm0.04$, the uncertainty
being the standard deviation of the five ratios divided by $\sqrt{5}$. The same chain with the PH-derived curves gives
4.03, 3.18, 2.80, 3.29 and 3.04\% ($F=1.08\pm0.07$). At 3\,GeV both chains give a better resolution than the PH (3.3 and
3.2\% against 3.6\%): the energy-to-energy scatter comes from the drift correction derived from the ToT
(section~\ref{sec:oreo}) rather than from the amplitude estimator, and the two values of $F$ are compatible with each
other and with the one obtained with the PH-based calibration ($F=1.07$).

A free-running dual-comparator readout can therefore be calibrated with a threshold scan and MIPs alone, without ever
digitizing the signals, at the same cost in resolution as with curves derived from the PH. This closes the loop for a
calorimeter readout without a digitizer.

\subsection{Lead glass}
\label{sec:lg}

\begin{figure}[htbp]
\centering
\includegraphics[width=\textwidth]{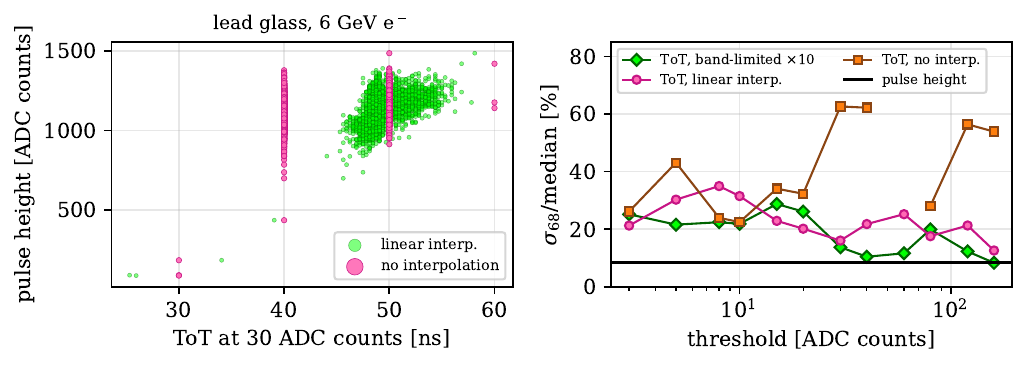}
\caption{Left: PH versus ToT at 30 ADC counts for the central LG block with 6\,GeV electrons, with and without
interpolation. Right: $\sigma_{68}$/median of the LG energy at 6\,GeV as a function of the threshold for the three ToT
flavors (points where all events share the same ToT value are not shown) and for the PH.}
\label{fig:lg}
\end{figure}

The LG behaves differently (figure~\ref{fig:lg}). Without interpolation, at 6\,GeV and a threshold of 30 ADC
counts the ToT of the central block takes essentially two values, 40 and 50\,ns, over the whole energy peak, and it
carries no information on the energy. With linear interpolation, or even after a band-limited ($\times10$ FFT) upsampling
of the waveform, the ToT is only weakly correlated with the PH, and $\sigma_{68}$/median ranges from 8\% to more than 30\%
depending on the threshold, against 8.3\% for the PH. The upsampled ToT reaches the PH value only at 160 ADC counts, where the
crossings fall on the few samples around the maximum that also determine the PH, and it changes erratically with the
threshold. The pulse, 1.7 samples wide at half maximum, contains frequencies
above the 50\,MHz Nyquist limit of the V1724, so that its shape between the samples, which determines the threshold
crossings, cannot be reconstructed. In addition, for a fast pulse without a long tail the ToT grows only as
$\sqrt{\ln(A/\theta)}$, which amplifies any timing uncertainty. These data do not allow the ToT of the LG to be
assessed: a sampling of 100\,MS/s is not adequate to emulate it.

% =============================================================================================
\section{Discussion and conclusions}
\label{sec:concl}

A dual-threshold ToT can replace the waveform digitizer for the energy measurement of OREO at the cost of about 8\%
in energy resolution, provided that the temperature dependence of the SiPM gain is controlled.
We compared, on the same events and with the same reconstruction chain, the energy resolution obtained with the pulse
height and with an offline emulation of the time over threshold for the OREO PbWO$_4$ calorimeter read out by SiPMs.
The main findings are:
\begin{itemize}
\item a single threshold cannot cover the dynamic range of the calorimeter: low thresholds are needed for the
outer crystals, but on the long SiPM tail of the central crystal a low threshold is either truncated by the acquisition
window or, without window, affected by the noise; the best single threshold worsens the resolution by 50--80\%;
\item a dual-threshold scheme, which works like a dual-gain front end with the two thresholds in place of the two
gains, recovers most of the performance with thresholds at about 0.6 and 6 MIPs: $\sigma/\mu=3.9\%$ at 2\,GeV and
$2.6\%$ at 6\,GeV,
with a core width 6--10\% and $\sigma_{68}$ 5--8\% larger than with the PH, comparable to the at most 8--12\% found for a
PET detector~\cite{Gaudin:2020dualToT}. The thresholds must be chosen looking at both the core
and the tails of the spectrum, since a gap between the two thresholds produces tails that a fit of the core does not see;
\item removing the acquisition window, emulated by extending the waveforms with the measured exponential tail and
noise, does not change these conclusions: a single threshold worsens the resolution by about 60\%, two thresholds by
7--8\%, and the choice of the two thresholds becomes much less critical. A comparator with hysteresis is not better than
a single threshold, and three or four thresholds reduce the degradation only to 4--6\%;
\item an analog shaper that shortens the SiPM tail to 50--100\,ns, followed by a 10\,ns low-pass filter, keeps the
dual-threshold degradation at 5--7\% and shortens the ToT from \SI{1.2}{\micro\second} to 0.3--0.5\,\si{\micro\second};
\item deriving all calibrations (pedestals, MIP equalization, drift correction) from the ToT alone keeps the
average degradation at 9\%, but makes the result less stable from one energy to the next, because the temperature
drift is corrected less precisely;
\item the ToT-to-amplitude conversion can itself be derived without a digitizer, from a scan of the two thresholds:
since the pulse shape depends only weakly on the amplitude, the ToT depends essentially only on the ratio between
threshold and amplitude, once an amplitude-independent shortening of the ToT caused by the noise at low thresholds is included.
With no information from the digitizer the degradation remains within the energy-to-energy fluctuations of the
ToT-based drift correction ($F=1.03\pm0.04$);
\item with 2\,ns sampling the interpolation of the threshold crossings brings no appreciable improvement, so that a
TDC with 2\,ns bins is sufficient, whereas with
10\,ns sampling and pulses less than two samples wide at half maximum (lead glass) the ToT cannot be emulated, with or without
interpolation;
\item the PH reference reproduces the mean energies and, once the same normalization is adopted, approximately the
calibration constants of the previous analysis~\cite{Selmi2026}; subtracting the pedestals in the MIP-based drift normalization
improves the resolution by 0.1--1.6 percentage points.
\end{itemize}

The study has some limitations that should be addressed before designing a ToT readout. The ToT-to-amplitude curves
of the main comparison are derived from the PH; the self-calibration of section~\ref{sec:nodigitizer} removes this
need, but the threshold scan is emulated on the recorded events, with the noise of the digitizer and exactly known
threshold values. With a ToT readout, the temperature dependence of the SiPM gain (about 10\%/$^\circ$C here) is much
harder to compensate (section~\ref{sec:oreo}): the MIP peak
used for the correction (about 25 ADC counts, \SI{3}{\milli\volt}) lies close to the low threshold, where the ToT is
least precise, and a gain change moves the signals with respect to the fixed thresholds, changing which signals are
lost below threshold and where the switch between the two thresholds occurs. A temperature-stabilized setup, or a bias
voltage adjusted with the temperature, would be needed.

The temperature drift also inflates the PH resolution itself: on single runs, each taken over at most two hours at
nearly constant temperature, the PH resolution at 3--5\,GeV is 10--25\% better than on the merged runs (e.g.\ 2.3\%
instead of 2.8\% at 5\,GeV), while the resolution with the ToT-only calibration remains 7--13\% worse than the PH
resolution. A stable
temperature is expected to improve the absolute resolution of OREO and to make the ToT-only calibration as reliable as the PH-based
one, but the ToT would still be 7--13\% worse than the PH. Finally, the noise is the one measured by the digitizer: the
comparator noise and bandwidth are not emulated, and the thresholds are optimized on the same data used to
measure the resolution. The window-less results rely on an extrapolation of the waveforms, validated on contained
pulses but without the AC coupling, and the shaper is emulated on the digitized signals.

\acknowledgments
We thank the CERN PS team for the beam operation, especially Dipanwita Banerjee for the substantial help and the
patience.

\bibliographystyle{JHEP}
\bibliography{refs}

\providecommand{\href}[2]{#2}\begingroup\raggedright\begin{thebibliography}{10}

\bibitem{Karpushkin2024}
N.~Karpushkin, D.~Finogeev, F.~Guber et~al., \emph{Analytical description of
  the time-over-threshold method based on time properties of plastic
  scintillators equipped with silicon photomultipliers}, {\emph{Nucl. Instrum.
  Meth. A} {\bfseries 1068} (2024) 169739}
  [\href{https://arxiv.org/abs/2308.08341}{{\ttfamily 2308.08341}}].

\bibitem{PenaRodriguez2025}
J.~{Pe\~na-Rodr\'iguez}, J.~F\"ortsch, C.~Pauly and K.-H. Kampert,
  \emph{Shaping circuit for improving linearity, bandwidth, and dynamic range
  in {ToT}-based detectors}, {\emph{Nucl. Instrum. Meth. A} {\bfseries 1084}
  (2026) 171273} [\href{https://arxiv.org/abs/2508.00739}{{\ttfamily
  2508.00739}}].

\bibitem{Shimazoe:2012dtot}
K.~Shimazoe, H.~Takahashi, B.~Shi, T.~Orita, T.~Furumiya, J.~Ooi et~al.,
  \emph{Dynamic time over threshold method},
  \href{https://doi.org/10.1109/TNS.2012.2215338}{\emph{IEEE Trans. Nucl. Sci.}
  {\bfseries 59} (2012) 3213}.

\bibitem{Fujiwara:2010mltot}
T.~Fujiwara, H.~Takahashi, K.~Shimazoe and B.~Shi, \emph{Multi-level
  time-over-threshold method for energy resolving multi-channel systems},
  \href{https://doi.org/10.1109/TNS.2010.2061236}{\emph{IEEE Trans. Nucl. Sci.}
  {\bfseries 57} (2010) 2545}.

\bibitem{Gaudin:2020dualToT}
{\'E}.~Gaudin, C.~Thibaudeau, L.~Arpin, H.~Bouziri, R.~Fontaine and R.~Lecomte,
  \emph{Dual-threshold {Time-over-Threshold} nonlinearity correction for {PET}
  detectors}, \href{https://doi.org/10.1016/j.nima.2020.164100}{\emph{Nucl.
  Instrum. Meth. A} {\bfseries 971} (2020) 164100}.

\bibitem{Thienpont:2020wau}
D.~Thienpont and C.~de~La~Taille, \emph{Performance study of {HGCROC-v2}: the
  front-end electronics for the {CMS} {High Granularity Calorimeter}},
  \href{https://doi.org/10.1088/1748-0221/15/04/C04055}{\emph{JINST} {\bfseries
  15} (2020) C04055}.

\bibitem{Bandiera2018}
L.~Bandiera, V.~V. Tikhomirov, M.~Romagnoni et~al., \emph{Strong reduction of
  the effective radiation length in an axially oriented scintillator crystal},
  {\emph{Phys. Rev. Lett.} {\bfseries 121} (2018) 021603}
  [\href{https://arxiv.org/abs/1803.10005}{{\ttfamily 1803.10005}}].

\bibitem{Soldani2025}
M.~Soldani, P.~Monti-Guarnieri, A.~Selmi et~al., \emph{Strong enhancement of
  electromagnetic shower development in oriented scintillating crystals and
  implications for particle detectors}, {\emph{Eur. Phys. J. C} {\bfseries 85}
  (2025) 1239} [\href{https://arxiv.org/abs/2404.12016}{{\ttfamily
  2404.12016}}].

\bibitem{Bandiera2025}
L.~Bandiera, P.~Fedeli and {the Oriented Calorimeter Collaboration},
  \emph{High-performance electromagnetic calorimeter with oriented crystals to
  open new pathways in particle and astroparticle physics},
  \href{https://doi.org/10.3389/fsens.2025.1659893}{\emph{Front. Sens.}
  {\bfseries 6} (2025) 1659893}.

\bibitem{Malagutti2024}
L.~Malagutti, A.~Selmi, L.~Bandiera et~al., \emph{High-precision alignment
  techniques for realizing an ultracompact electromagnetic calorimeter using
  oriented high-{Z} scintillator crystals}, {\emph{Nucl. Instrum. Meth. A}
  {\bfseries 1069} (2024) 169869}
  [\href{https://arxiv.org/abs/2503.17053}{{\ttfamily 2503.17053}}].

\bibitem{Selmi2026}
A.~Selmi, \emph{Oriented crystals: an innovative approach to electromagnetic
  calorimetry}, Ph.D. thesis, Universit\`a degli Studi dell'Insubria,
  \url{https://tesidottorato.depositolegale.it/bitstream/20.500.14242/363106/1/PHDThesisSelmi.pdf},
  2026.

\bibitem{Prest2003}
M.~Prest, G.~Barbiellini, G.~Bordignon, G.~Fedel, F.~Liello, F.~Longo et~al.,
  \emph{The {AGILE} silicon tracker: an innovative $\gamma$-ray instrument for
  space}, \href{https://doi.org/10.1016/S0168-9002(02)02047-8}{\emph{Nucl.
  Instrum. Meth. A} {\bfseries 501} (2003) 280}.

\bibitem{HamamatsuS13360}
{Hamamatsu Photonics}, ``{MPPC} (multi-pixel photon counter) {S13360} series,
  datasheet {KAPD1052E}.''
  \url{https://www.hamamatsu.com/content/dam/hamamatsu-photonics/sites/documents/99_SALES_LIBRARY/ssd/s13360_series_kapd1052e.pdf},
  2024.

\bibitem{Das2016}
S.~Das, \emph{A simple alternative to the {Crystal Ball} function},
  \href{https://arxiv.org/abs/1603.08591}{{\ttfamily 1603.08591}}.

\end{thebibliography}\endgroup

\end{document}